\documentclass[aps,prc,twocolumn,superscriptaddress,floatfix,showpacs,nofootinbib]{revtex4-1}

\usepackage{graphicx}
\usepackage{graphics}
\usepackage{subfigure}
\usepackage{here}
\usepackage[utf8]{inputenc}
\usepackage[T1]{fontenc}
\usepackage{amssymb} 
\usepackage{amsmath} 
\usepackage{color}

\usepackage{siunitx}
\usepackage{lmodern}
\usepackage{textcomp}

\newcommand{\ivs}{$\rm{s}^{-1}$}

\newcommand{\lmup}{\lambda^{}_{+}}
\newcommand{\lmum}{\lambda^{}_{-}}

\newcommand{\LpAr}{\Lambda^{}_{p\rm{Ar}}}
\newcommand{\LpZ}{\Lambda^{}_{p\rm{Z}}}
\newcommand{\lpAr}{\lambda^{}_{p\rm{Ar}}}
\newcommand{\lpZ}{\lambda^{}_{p\rm{Z}}}
\newcommand{\LAr}{\Lambda^{}_{\rm{Ar}}}
\newcommand{\LZ}{\Lambda^{}_{\rm{Z}}}
\newcommand{\Lppmu}{\Lambda^{}_{pp\mu}}
\newcommand{\lppmu}{\lambda^{}_{pp\mu}}
\newcommand{\Lpf}{\Lambda^{}_{\mathrm{pf}}}
\newcommand{\lpf}{\lambda^{}_{\mathrm{pf}}}
\newcommand{\Lof}{\Lambda^{}_{\mathrm{of}}}
\newcommand{\lof}{\lambda^{}_{\mathrm{of}}}

\newcommand{\nmup}{n^{}_{p\mu }}
\newcommand{\nmuAr}{n^{}_{\rm{Ar}\mu}}
\newcommand{\nmuZ}{n^{}_{\rm{Z}\mu}}

\newcommand{\northo}{n^{}_{\rm{om}}}
\newcommand{\npara}{n^{}_{\rm{pm}}}

\newcommand{\nprimemup}{n'_{p\mu }}

\newcommand{\nprimemuZ}{n'_{\rm{Z}\mu}}

\newcommand{\nprimeortho}{n'_{\rm{om}}}
\newcommand{\nprimepara}{n'_{\rm{pm}}}

\newcommand{\LS}{\Lambda^{}_{\rm{S}}}
\newcommand{\LT}{\Lambda^{}_{\rm{T}}}
\newcommand{\lop}{\lambda^{}_{\rm{op}}}
\newcommand{\LO}{\Lambda^{}_{\rm{OM}}}
\newcommand{\LP}{\Lambda^{}_{\rm{PM}}}
\newcommand{\den}{\phi}

\newcommand{\pmuS}{$(p\mu)^{}_{\rm{S}}$}
\newcommand{\ppmuO}{$(pp\mu)^{}_{\rm{om}}$}
\newcommand{\ppmuP}{$(pp\mu)^{}_{\rm{pm}}$}

\newcommand{\Gpmu}{\Gamma^{}_{p\mu}}
\newcommand{\Gom}{\Gamma^{}_{\rm{om}}}
\newcommand{\Gpm}{\Gamma^{}_{\rm{pm}}}
\newcommand{\Gz}{\Gamma^{}_{\rm{Z}\mu}}
\newcommand{\GAr}{\Gamma^{}_{\rm{Ar}\mu}}

\usepackage{epsfig}
\usepackage{bbold}
\usepackage{comment}

\newcommand\T{\rule{0pt}{2.6ex}}
\newcommand\B{\rule[-1.2ex]{0pt}{0pt}} 

\newcommand{\apsi}
{\affiliation{Paul Scherrer Institute, CH-5232 Villigen PSI, Switzerland}}
\newcommand{\aucb}
{\affiliation{Department of Physics, University of California, Berkeley, and LBNL, Berkeley, CA 94720, USA}}
\newcommand{\apnpi}
{\affiliation{Petersburg Nuclear Physics Institute, Gatchina 188350, Russia}}
\newcommand{\auiuc}
{\affiliation{Department of Physics, University of Illinois at Urbana-Champaign, Urbana, IL 61801, USA}}
\newcommand{\auw}
{\affiliation{Department of Physics, University of Washington, Seattle, WA 98195, USA}}
\newcommand{\aucl}
{\affiliation{Institute of Nuclear Physics, Universit{\'e} Catholique de Louvain, B-1348, Louvain-la-Neuve, Belgium}}
\newcommand{\auk}{\affiliation{Department of Physics and Astronomy, University of Kentucky, Lexington, KY 40506, USA}}
\newcommand{\aub}{\affiliation{Department of Physics, Boston University, Boston, MA 02215, USA}}
\newcommand{\aregis}
{\affiliation{Department of Physics and Computational Science, Regis University, Denver, CO 80221, USA}}
\newcommand{\ayork}
{\affiliation{Department of Earth and Physical Sciences, York College,
  City University of New York, Jamaica, NY 11451, USA}}

\newcommand{\mathis}{\textrm{s}^{-1}}

\newcommand{\gpm}{g^{}_{P}}

\newcommand{\LambdaMuplus}{455170.05}
\newcommand{\LambdaMuplusErr}{0.46}

\begin{document}

\title{Measurement of the Formation Rate of Muonic Hydrogen Molecules}
\author{V.A.~Andreev}
\apnpi
\author{T.I.~Banks}
\aucb
\author{R.M.~Carey}
\aub
\author{T.A.~Case}
\aucb
\author{S.M.~Clayton}
\thanks{Present address: Los Alamos National Laboratory, Los Alamos, NM 87545, USA}
\auiuc
\author{K.M.~Crowe}\thanks{Deceased}
\aucb
\author{J.~Deutsch}\thanks{Deceased}
\aucl
\author{J.~Egger}
\apsi
\author{S.J.~Freedman}\thanks{Deceased}
\aucb
\author{V.A.~Ganzha}
\apnpi
\author{T.~Gorringe}
\auk
\author{F.E.~Gray}
\aregis
\aucb
\author{D.W.~Hertzog}
\auw\auiuc
\author{M.~Hildebrandt}
\apsi
\author{P.~Kammel}
\auw\auiuc
\author{B.~Kiburg}\thanks{Present address: Fermi National Accelerator Laboratory, Batavia, IL 60510, USA}
\auw\auiuc
\author{S.~Knaack}
\thanks{Present address: University of Wisconsin, Madison, WI 53706, USA}
\auiuc
\author{P.A.~Kravtsov}
\apnpi
\author{A.G.~Krivshich}
\apnpi
\author{B.~Lauss}
\apsi
\author{K.R.~Lynch}
\ayork
\author{E.M.~Maev}
\apnpi
\author{O.E.~Maev}
\apnpi
\author{F.~Mulhauser}
\auiuc
\apsi
\author{C.~Petitjean}
\apsi
\author{G.E.~Petrov}
\apnpi
\author{R.~Prieels}
\aucl
\author{G.N.~Schapkin}
\apnpi
\author{G.G.~Semenchuk}
\apnpi
\author{M.A.~Soroka}
\apnpi
\author{V.~Tishchenko}
\thanks{Present address: Brookhaven National Laboratory, Upton, NY 11973, USA}
\auk
\author{A.A.~Vasilyev}
\apnpi
\author{A.A.~Vorobyov}
\apnpi
\author{M.E.~Vznuzdaev} 
\apnpi
\author{P.~Winter}\thanks{Corresponding author}
\email{winterp@anl.gov}
\thanks{Present address: Argonne National Laboratory, Lemont, IL 60439, USA}
\auw\auiuc
\collaboration{MuCap Collaboration}
\noaffiliation
\date{\today}

\begin{abstract}
\begin{description}
\item[Background] The rate $\lppmu$ characterizes the formation of $pp\mu$ molecules in collisions of muonic $p\mu$ atoms with hydrogen.  In measurements of the basic weak muon capture reaction on the proton to determine the pseudoscalar coupling $\gpm$, capture occurs from both atomic and molecular states. Thus knowledge of $\lppmu$ is required for a correct interpretation of these experiments. 
\item[Purpose] Recently the MuCap experiment has measured the capture
  rate $\LS$ from the  singlet  $p\mu$ atom, employing a low
  density active target to suppress $pp\mu$ formation (PRL 110, 12504 (2013)). Nevertheless, 
given the unprecedented precision of this experiment, the existing experimental knowledge in $\lppmu$ had to be improved.
\item[Method] The MuCap experiment derived the weak capture rate from the muon disappearance rate in ultra-pure hydrogen. By doping the hydrogen with 20 ppm of argon, a competing process to $pp\mu$
formation was introduced, which allowed the extraction of $\lppmu$ from the observed time distribution of decay electrons.
\item[Results] 
The  $pp\mu$ formation rate was measured as $\lppmu=\left(2.01 \pm
  0.06_{\mathrm{stat}}\pm
  0.03_{\mathrm{sys}}\right)\times10^6~\mathrm{s}^{-1}$. This result
updates the $\lppmu$ value used in the above mentioned MuCap
publication. 
\item[Conclusions] 
The 2.5$\times$ higher precision compared to earlier experiments and
the fact that the measurement was performed at nearly identical
conditions to the main data taking, reduces the uncertainty induced by
$\lppmu$ to a minor contribution to the overall uncertainty of $\LS$ and $\gpm$, as determined in MuCap. 
Our final value for $\lppmu$  shifts  $\LS$ and $\gpm$ 
by less than one tenth of their respective uncertainties compared to
our results published earlier.  
\end{description}
\end{abstract}
\maketitle

\section{Introduction}
\label{Introduction}
Nuclear muon capture on the proton,
\begin{equation}
\mu^{-} + p \rightarrow n + \nu_{\mu} ~, \label{equation:omc}
\end{equation}
is a basic charged-current weak reaction~\cite{Czarnecki:2007th, Kammel:2010zz, Gorringe:2002xx}. Several experiments have measured the rate of ordinary muon capture (Eq.~\eqref{equation:omc}) or the rarer process of radiative muon capture, $\mu + p \rightarrow n + \nu +\gamma$, in order to determine the weak pseudoscalar coupling of the proton, $\gpm$, which can be extracted most straightforwardly from muon capture on the nucleon.
A precision determination of $\gpm$ has been a longstanding experimental challenge~\cite{Gorringe:2002xx,Kammel:2010zz} due to the small rate of capture on the proton and complications arising from the formation of muonic molecules. The most recent MuCap result, $\gpm = 8.06 \pm 0.55$~\cite{Andreev:2012fj}, achieved an unprecedented precision of 7\,\%, thereby providing a sensitive test of QCD symmetries and confirming a fundamental prediction of chiral perturbation theory, $\gpm = 8.26 \pm 0.23$~\cite{Bernard:1994wn, Bernard:2001rs, Pastore:2014iga}.

Experimentally, process~(\ref{equation:omc}) is observed after low-energy muons are stopped in hydrogen, where they form
$p\mu$ atoms and $pp\mu$ molecules. The overlap in the wavefunctions of the proton and the bound muon leads to small but observable capture rates at the $10^{-3}$ level relative to muon decay, $\mu^- \rightarrow e^- \bar{\nu}_e \nu_\mu$, which is the dominant mode of muon disappearance in that environment. The nuclear capture rates depend on the spin compositions of the muonic atoms and molecules (a direct consequence of the $\rm{V}\!\!-\!\!\rm{A}$ structure of the electroweak interaction), and thus the rates vary significantly among the different muonic states. The calculated rates for the two hyperfine states of the $p\mu$ atom possessing spin F=0,1 are $\LS=712.7$ \ivs\ and $\LT=12.0$ \ivs, respectively (c.f.~\cite{Czarnecki:2007th}, updated in~\cite{Andreev:2012fj}). The formation of $pp\mu$ molecules further complicates the situation, as the calculated capture rates for the ortho and para states, $\LO$= 542.4 \ivs\ and $\LP$= 213.9 \ivs, differ too from the atomic rates (Eq.~\eqref{Rop.eq}). Correct interpretation of the observed muon disappearance rate thus relies on a thorough understanding of the ``muon chemistry'' reactions governing the time evolution of the $p\mu$ and $pp\mu$ states. This interrelationship between muon capture and muon chemistry in hydrogen has been the primary source of ambiguity in the 50-year history of experiments in the field. 
Historically, interest in muon atomic and molecular reactions arose due to their above mentioned relevance for the determination of nuclear muon capture rates in hydrogen isotopes~\cite{Kammel:2010zz} and their importance in muon-catalyzed fusion~\cite{Breunlich:1989vg}, where $\lppmu$ was calculated within a systematic program to solve the
Coulomb three-body problem~\cite{Faifman:1989}.

The MuCap experiment employed a novel technique involving the use of low-pressure hydrogen gas to suppress molecular formation. Nevertheless, it was still necessary to apply corrections which were based on measurements of the molecular formation rates that determine the $pp\mu$ ortho and para molecule populations. In the initial MuCap physics result~\cite{Andreev:2007wg}, we conservatively estimated that the uncertainty in the molecular formation rate $\lppmu$ contributed a systematic uncertainty of 4.3\,\ivs\ to our determination of $\LS$, the muon capture rate in the $p\mu$ hyperfine singlet state. 
During the later high statistics data taking for MuCap, we performed a dedicated measurement of $\lppmu$ in order to improve the precision on this parameter and 
render its contribution to the uncertainty on $\LS$ nearly negligible. The final 
MuCap result, $\LS = (714.9\pm5.4_{\rm{stat}}\pm 5.1_{\rm{syst}})$\,\ivs\ \cite{Andreev:2012fj}, possessed greatly improved statistical and systematic uncertainties. 
A preliminary value for $\lppmu$ obtained from our measurement was an important ingredient to this result. In this paper we document the $\lppmu$ experiment and present its final results.

The contents of this article are as follows: In Section~\ref{CaptureChemistry} we introduce muon-induced processes in hydrogen and their impact on muon-capture measurements. In Section~\ref{MuCap} we describe the MuCap experiment and our technique for measuring $\lppmu$; the corresponding data analysis and result for $\lppmu$ are described in Section~\ref{Experiment_ppm}. In Section~\ref{Interpretation} we use our new result for $\lppmu$ to update previous MuCap measurements. A concluding summary is given in Section~\ref{Summary}.

\section{Muon Capture and Muon Chemistry}
\label{CaptureChemistry}
\subsection{Muon Reactions in Hydrogen\label{sec:muonsinhydrogen}}
\begin{figure}[htb]
{\centering \includegraphics[width=0.99\linewidth]{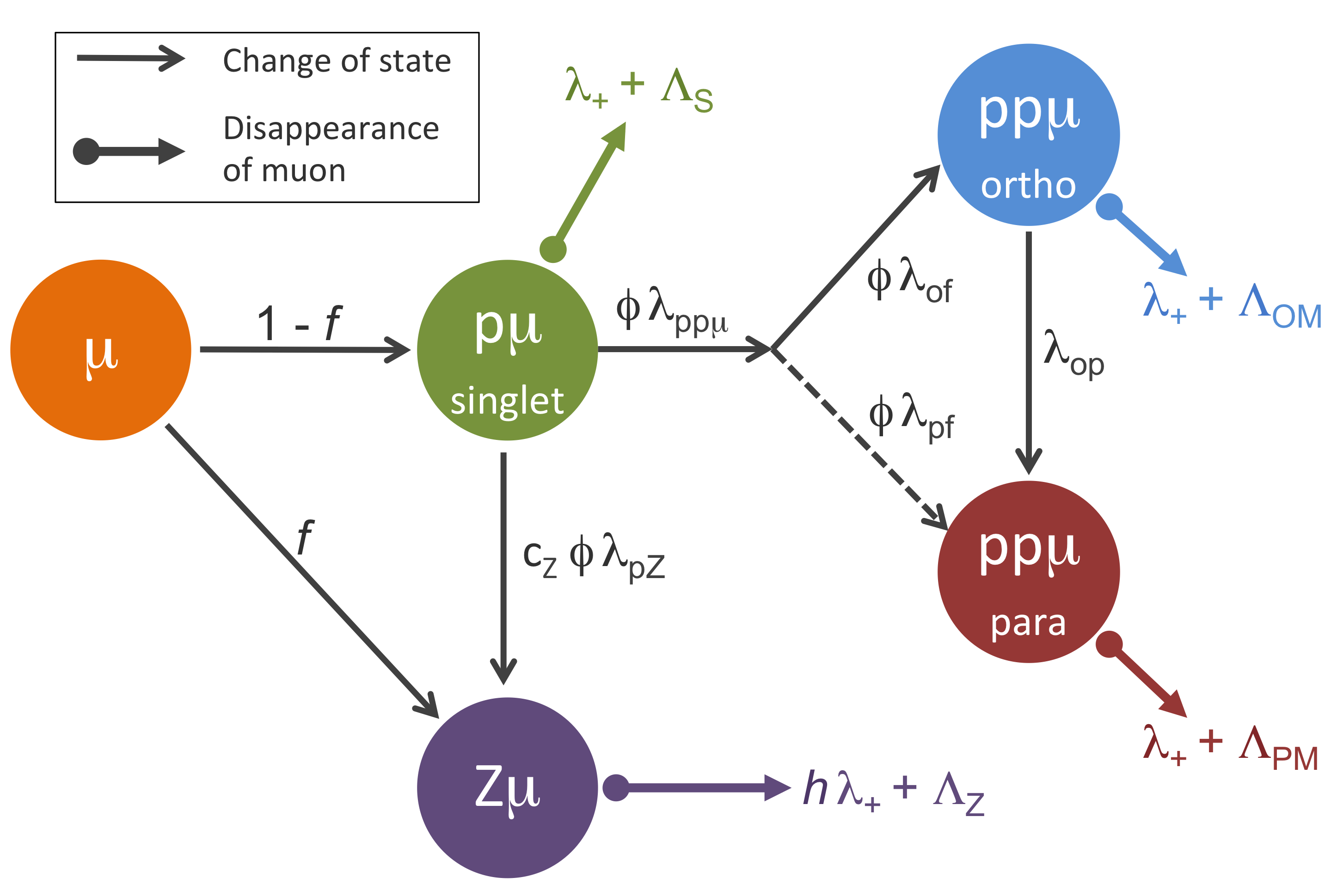}}
    \caption{(Color online) Kinetics of negative muons in hydrogen containing a single 
    	 chemical impurity Z. The large circles represent the different muonic states 
	 that can form. The black arrows denote transitions between muonic states, while
	 the colored arrows with a small circle at the beginning of the line
      indicate muon disappearance due to the weak-interaction processes of decay
      or nuclear capture. The arrow labeled $\lpf$ is dashed to indicate that its rate
      is about 240~times smaller than $\lof$.
}\label{FullModel}
\end{figure}

Muons stopped in hydrogen can form a variety of atomic and molecular states which are 
subject to different physical processes and whose populations are governed by the rates 
shown in Fig.~\ref{FullModel}. Table~\ref{tab:relevantrates} lists all of the rates used in this 
paper and their values. Several of the atomic processes proceed via binary collisions of 
muonic atoms with other target molecules. It is conventional to normalize those density-dependent
rates to the values observed at LH$_2$ density, $\den^{}_0= 4.25\times
10^{22}$\,atoms/cm$^3$, and express all target densities $\den$ relative to $\den^{}_0$. 

\begin{table}[htb]
\caption{Summary of relevant rates for processes involving muons in hydrogen.\label{tab:relevantrates}}
{\centering 
\begin{tabular}{llrl}
\hline
\hline
\T \B Muon process & Symbol & Value [\ivs] & Reference\\
\hline
\multicolumn{4}{l}{\T Weak-interaction rates}\\
Muon decay & $\lmup$ & $\LambdaMuplus \pm \LambdaMuplusErr$ & \cite{Agashe:2014kda, Tishchenko:2012ie, Webber:2010zf}\\ 
$p\mu$ singlet capture & $\LS$ & $714.9\pm 7.4$ & \cite{Andreev:2012fj}\\
$p\mu$ triplet capture & $\LT$ & $12.0 \pm 0.1$  & \cite{Czarnecki:2007th}\\
$pp\mu$ ortho capture & $\LO$ & $544.0 \pm 11.3$ & Eq.~\eqref{Rop.eq}\\
$pp\mu$ para capture & $\LP$ & $214.6 \pm 4.2$ & Eq.~\eqref{Rop.eq}\\
N capture & $\Lambda^{}_{\rm{N}}$ & $6.93 \pm 0.08 \times 10^4$ & \cite{Suzuki:1987jf}\\
O capture & $\Lambda^{}_{\rm{O}}$ & $10.26 \pm 0.06 \times 10^4$ & \cite{Suzuki:1987jf}\\
Ar capture & $\LAr$ & $141 \pm 11 \times 10^4$ &\cite{Suzuki:1987jf}\\
\B                       &&$\mathbf{130.2 \pm 3.2 \times 10^4}$
                       &  \textbf{this work}\\
\hline
\multicolumn{4}{l}{\T Atomic and molecular rates}\\
o-p transition & $\lop$ & $6.6 \pm 3.4 \times 10^4$ & \cite{Kammel:2010zz}\\
$pp\mu$ formation\footnote[1]{normalized to LH$_2$ density $\den^{}_0$} & $\lppmu$& $2.3 \pm 0.2  \times 10^6$ & \cite{Kammel:2010zz}\\
                      &&$\mathbf{2.01  \pm  0.07  \times
                        10^6}$
                       &  \textbf{this work}\\
Transfer to N\footnotemark[1] & $\lambda_{p\rm{N}}$ & 0.34 $ \pm 0.07 \times 10^{11}$ & \cite{Thalmann1998}\\
Transfer to O\footnotemark[1] & $\lambda_{p\rm{O}}$ & 0.85 $ \pm 0.02 \times 10^{11}$ & \cite{ON_Werthmuller}\\
Transfer to Ar\footnotemark[1]& $\lpAr$ & $1.63\pm 0.09 \times 10^{11}$ & \cite{PhysRevA.55.3447}\\
\B                       &&$\mathbf{1.94 \pm 0.11 \times 10^{11} }$
                       &  \textbf{this work}\\
\hline
\T \B Ar Huff factor\footnote[2]{dimensionless quantity} & $h$ & $0.985 \pm 0.003$ &\cite{ABSE_Huff,Watanabe1993165,Watanabe:1987su}\\
\hline
\hline
\end{tabular}}
\end{table}

Negatively charged, low-energy muons entering hydrogen are slowed down and 
undergo atomic capture, forming highly excited $p\mu$ atoms. 
After an atomic cascade to the ground state, the two hyperfine states of the 
$p\mu$ atom, singlet (F=0) and triplet (F=1), are populated according to their
statistical weights $\frac{1}{4}$ and $\frac{3}{4}$, respectively. 
These complex initial stages happen on a timescale of nanoseconds
at target densities exceeding $\den \ge$ 0.01, as in our case. 
Charge-exchange collisions~\cite{Gershtein:1958, Gershtein:1961}
convert the higher-lying triplet state to the lower-lying singlet state at
a rate calculated to be $\approx  \den \times 2 \cdot
10^{10}$\;\ivs~\cite{PhysRevA.43.4668}. Thus after less than 100~ns 
the triplet state is effectively depopulated and the main
features of the kinetics can be described by the scheme depicted in
Fig.~\ref{FullModel}. This condition is true for the present analysis and 
in Refs.~\cite{Andreev:2007wg, Andreev:2012fj}.

For our purposes, muon kinetics in pure hydrogen effectively starts
with the $p\mu$ atom in its hyperfine singlet state. 
In subsequent collisions of the $p\mu$ atom with hydrogen molecules, two types of $pp\mu$ molecules can be formed which differ in their angular momentum~L and total spin~I. Due to the Fermi statistics of the two-proton system, the ortho state \ppmuO\ has L=1, I=1, while the para state \ppmuP\ has L=0, I=0.
According to theory, $pp\mu$ formation proceeds to the ortho state predominantly at the normalized rate
$\lof=1.8\times 10^6$ \ivs, 
while the para formation rate $\lpf$=$7.5\times 10^3$ \ivs\ is much
smaller~\cite{Faifman:1989}. The total normalized molecular formation rate is the sum of these two rates,
\begin{equation}
\label{eq:molecularformation}
\lppmu = \lof + \lpf ~.
\end{equation}
Molecular formation scales with the target density $\phi$, so experiments observe the effective molecular formation rate
\begin{equation}
\Lppmu = \phi\lppmu ~.
\label{eq:effectiverate}
\end{equation}

The transition from the $pp\mu$ ortho state to the lower para state at rate $\lop$ involves a proton spin
flip and is only allowed due to relativistic effects in the molecular
wave function. The  $(pp\mu_{\rm{om}})^+$ is positively charged and
quickly forms various molecular complexes in collisions with H$_2$ molecules. 
The ortho-para transition proceeds at the calculated
rate $\lop=(7.1 \pm 1.2) \times 10^4$\,\ivs\ via the emission of an 
electron from these clusters~\cite{Bakalov:1980fm}. Two previous experiments
measured $\lop$ and obtained the inconsistent results
$(4.1 \pm 1.4) \times 10^4$\,\ivs~\cite{Bardin:1981cq}
and $(11.1 \pm 1.9) \times 10^4$\,\ivs~\cite{Clark:2005as}.
Review~\cite{Kammel:2010zz} therefore inflated the uncertainties and quoted an
average experimental value of $\lop= (6.6 \pm 3.4) \times 10^4$\,\ivs, 
which we use in this work. 

As mentioned above, the weak nuclear capture rates strongly depend on spin factors 
within the total $pp\mu$ molecular spin function and can be expressed as
\begin{equation}
\begin{split}
\LO &=2 \gamma^{}_{\rm{om}} (\tfrac{3}{4} \LS + \tfrac{1}{4} \LT), \\
\LP &=2 \gamma^{}_{\rm{pm}} (\tfrac{1}{4} \LS + \tfrac{3}{4} \LT).
\end{split}
 \label{Rop.eq}
\end{equation}
The molecular overlap factors are $2 \gamma^{}_{\rm{om}}=1.009\pm 0.001$ and
$2 \gamma^{}_{\rm{pm}}= 1.143\pm 0.001$~\cite{Bakalov:1980fm}. Based on these
equations the capture rates of the molecular states can be calculated
using the MuCap result for $\LS$ and the theoretical value for the
smaller rate $\LT$ as input (see Table~\ref{tab:relevantrates}). 
 
In the presence of $\rm{Z}>1$ chemical impurities, the muon can
form a bound Z$\mu$ state instead of a $p\mu$ atom. The
factor $f$ in Fig.~\ref{FullModel} characterizes the initial
population of Z$\mu$ atoms, which arises from two pathways. 
First, at the time of the muon stop, $\rm{Z}>1$ elements are energetically favored over hydrogen by Coulomb capture.
Second, during the $p\mu$ deexcitation cascade, prompt transfer to higher Z elements can occur. 
The size of $f$ scales linearly with the relative atomic concentration $c^{}_{\rm{Z}}$ of the impurity. 

Muons will also transfer from the singlet $p\mu$ state to the energetically favorable
Z$\mu$ state in collisional processes. Transfer from the molecular states to the Z$\mu$ state is not possible because the charged $(pp\mu)^+$ molecule is repelled by the Z~nucleus. The effective transfer rate to the impurity, $\LpZ$, is expressed as 
\begin{equation}
\Lambda^{}_{p\rm{Z}}=c^{}_{\rm{Z}} \phi \lpZ ~,
\label{eq:ImpurityTransfer}
\end{equation}
where $\lpZ$ is the normalized transfer rate. Excited Z$\mu$ states are created by such transfers, and observable muonic X-rays are emitted during the subsequent deexcitation cascade. The rate $\Lambda^{}_{\rm{Z}}$ of subsequent muon capture on the nucleus increases roughly proportional to Z$^{4}$ (The more realistic Primakoff formula is discussed in~\cite{Suzuki:1987jf}.). Table~\ref{tab:relevantrates} shows that the capture rates for typical impurity elements (nitrogen, oxygen, argon) are all much larger than the $p\mu$ singlet capture rate $\LS$.

The natural abundance of deuterium in hydrogen generally causes an additional loss channel due to the formation of $d\mu$ atoms \cite{Lauss:1996zz, Lauss:1999} and $pd\mu$ molecules \cite{Gorringe:2002xx}. For the presented measurement, a cryogenic distillation column was used to isotopically purify the hydrogen achieving a final deuterium concentration of less than 10\,ppb \cite{Andreev:2012fj}. At this level, the deuterium loss channel is completely negligible. 

The muon can decay from any of the states in Fig. \ref{FullModel} at a rate close to the free muon decay rate, $\lmup$~\cite{Agashe:2014kda, Webber:2010zf}. The actual decay rates are slightly reduced with respect to $\lmup$ by the Huff factor $h$~\cite{ABSE_Huff} which accounts for bound-state corrections arising from Coulomb and relativistic effects.
We neglect the Huff factor in the $p\mu$ system in the following equations, as it is calculated to reduce $\lmup$ by only
26\,ppm~\cite{Uberall:1960zz, VonBaeyer:1979yg}; in the final evaluation of $\LS$, Eq.~\eqref{eq:decayrate}, we will
explicitly include this reduction. 
For the argon system we use $h=0.985\pm 0.003$, based on extended-model calculations~\cite{Watanabe1993165,Watanabe:1987su} which include a more accurate treatment of finite nuclear size effects.

\subsection{Kinetic Equations}
The kinetics scheme in Fig.~\ref{FullModel} corresponds to a system of coupled linear differential equations for the time-dependent populations $\nmup(t)$, $\northo(t)$, $\npara(t)$ and $\nmuZ(t)$ of the $p\mu$, $pp\mu$, and Z$\mu$ states. 
It is convenient to first define the total muon disappearance rate from each state:
\begin{equation}
\begin{split}
\Gpmu&\equiv\lmup+\LpZ+\LS+\Lppmu\\
\Gom&\equiv\lmup+\lop+\LO\\
\Gpm&\equiv\lmup +\LP\\
\Gz&\equiv h\lmup+\LZ.
\end{split}\label{eq:eigen}
\end{equation}
These rates are also the eigenvalues of the system.
The populations of the muonic states are then described by the following differential equations:
\begin{equation}
\begin{split}
\nprimemup(t)&=-\Gpmu\;\nmup(t) \\
\nprimeortho(t)&=\Lof\;\nmup(t)-\Gom\;\northo(t)\\
\nprimepara(t)&=\Lpf\;\nmup(t)+\lop\;\northo(t)-\Gpm\;\npara(t)\\
\nprimemuZ(t)&=\LpZ\;\nmup(t)-\Gz\;\nmuZ(t).\label{eq:N}
\end{split}
\end{equation}
The initial conditions at $t=0$ are $\nmup(0)=1-f$, $\nmuZ(0)=f$ and
$\northo(0)=\npara(0)=0$. If the rates are time
independent, Eq.~\eqref{eq:N} defines a system of differential equations
with \textit{constant} coefficients which has straightforward but
lengthy analytical solutions $n_i(t)$ (given in the
appendix). Formally they can be written in terms of the 
eigenvalues $\Gamma^{}_\alpha$ (see Eqs.~\eqref{eq:eigen}):
\begin{equation}
n^{}_i(t)= \sum_\alpha c^{\alpha}_{i} \; e^{-\Gamma_\alpha t}_{} ~,
\label{eq:alpha}
\end{equation}
where $i, \alpha \in (p\mu,\ \rm{om,\ pm,\ }Z\mu)$.
Usually this is a good approximation, but as will be explained later there are cases where epithermal
$p\mu$ atoms are depopulated at energy-dependent rates in the period before they have fully 
thermalized (c.f.~\cite{Werthmueller:1996, Adamczak:2007}). 
Of particular relevance for the present work, a muonic X-ray measurement~\cite{PhysRevA.55.3447} 
observed that the muon transfer rate $\lambda_{p{\rm{Ar}}}(t)$ increased until it reached its constant value
for the thermalized atom. 
Because $\lambda_{p{\rm{Z}}}(t)$ is time dependent, 
Eq.~\eqref{eq:N} must be numerically integrated.

From the muonic state populations $n^{}_i(t)$ we can derive the time distributions 
of various experimentally observed final-state muon-disappearance products. 
The distribution of decay electrons is given by
\begin{equation}
\begin{split}
N_e(t)=\lmup\Big\{&\epsilon_e\;\big[\nmup(t)+\npara(t)+\northo(t)\big] + {}\\
&\epsilon'_e\; h\;\nmuZ(t)\Big\} ~,
\label{eq:Ne}
\end{split}
\end{equation}
where $\epsilon_e$ is the detection efficiency for electrons produced by muon 
decay from the hydrogen bound states. Depending on the experimental setup, 
the detection efficiency $\epsilon'_e $ in higher-Z atoms can be different because 
the electron energy spectrum deviates from a pure Michel spectrum due to 
Coulomb effects~\cite{Watanabe1993165,Watanabe:1987su}.

The distribution of muon capture products (i.e., recoil nuclei or neutrons) versus time is
\begin{eqnarray}
N_c(t)&=&\epsilon_c \big[\LS \nmup(t) +\LO \northo(t) + \LP \npara(t)\big]+
\nonumber\\
&&\epsilon'_c \LZ \nmuZ(t) ~.
\label{eq:Nc}
\end{eqnarray}
Here $\epsilon_c$ and $\epsilon'_c$ account for the different
efficiencies in detecting reaction products from capture on protons versus 
capture on nuclei with atomic number Z.

The time distribution of X-rays from muon transfer is
\begin{equation}
N_x(t)= \epsilon_x P_x \LpZ \nmup(t)  \label{eq:Nx} ~,
\end{equation} 
where $P_x$ is the probability for X-ray emission per transfer and
$\epsilon_x$ is the X-ray detection efficiency.  The observables in
Eqs.~\eqref{eq:Ne},~\eqref{eq:Nc} and~\eqref{eq:Nx} provide the 
primary tools for experimentalists in disentangling the rich physics 
of muon-induced processes in hydrogen.

\subsection{Present Experimental Knowledge of the Molecular Formation Rate $\lppmu$}

The basic experimental technique for measuring the molecular formation
rate $\lppmu$ is to introduce an impurity to the pure hydrogen target. 
Though it might seem counterintuitive, adding this complication is helpful
because it opens a competing channel to molecular $pp\mu$ formation. 
Since muon transfer to the impurity only proceeds from the $p\mu$ atom, the Z$\mu$
population follows the time evolution of the $p\mu$ population which feeds it and the 
electron distribution described in Eq.~\eqref{eq:Ne} depends mainly on $\Lppmu$, $\LpZ$, and $\LZ$. 
By adding the proper amount of a well-chosen impurity, the terms in Eq.~\eqref{eq:Ne} 
will differ in their time dependencies and relative sizes such that individual rates can be 
disentangled via a fit to the observed electron time spectrum.

An early measurement of $\lppmu$ used an LH$_2$ target with deuterium 
admixtures~\cite{PhysRev.132.2679}. In this case, muons transfer from $p\mu$
to $d\mu$ and deuterium essentially plays the role of the
impurity Z in Fig.~\ref{FullModel} and Eq.~\eqref{eq:Ne}. 
The formation of $d\mu$ atoms can lead to muon-catalyzed fusions which 
emit gammas. 
Observation of the gamma yields for various deuterium concentrations
thus enabled a determination of $\lppmu$.  

Other experiments employed a similar strategy. Ref.~\cite{Conforto:1964}
measured the muonic X-rays emitted following transfer to Ne.
Experiment~\cite{Bystritskii:1976}
simultaneously observed the time 
distribution of $\mu$Xe deexcitation X-rays and muon decay electrons. The first measurable determined 
$\Gpmu$, while the second enabled independent extraction of $\Lppmu$ and $\LpZ$ at a single impurity concentration.

The most recent experiment~\cite{Mulhauser:1996zz} used a very different experimental setup
consisting of a layer of solid hydrogen with various tritium admixtures.
Fusion products were observed, and muon transfer to tritium changed
the disappearance rate of the $p\mu$ state according to the first
of Eqs.~\eqref{eq:N}. Conceptually the experiment was therefore
quite similar to~\cite{PhysRev.132.2679}.

\begin{figure}[htb]
{\centering \includegraphics[width=0.99\linewidth]{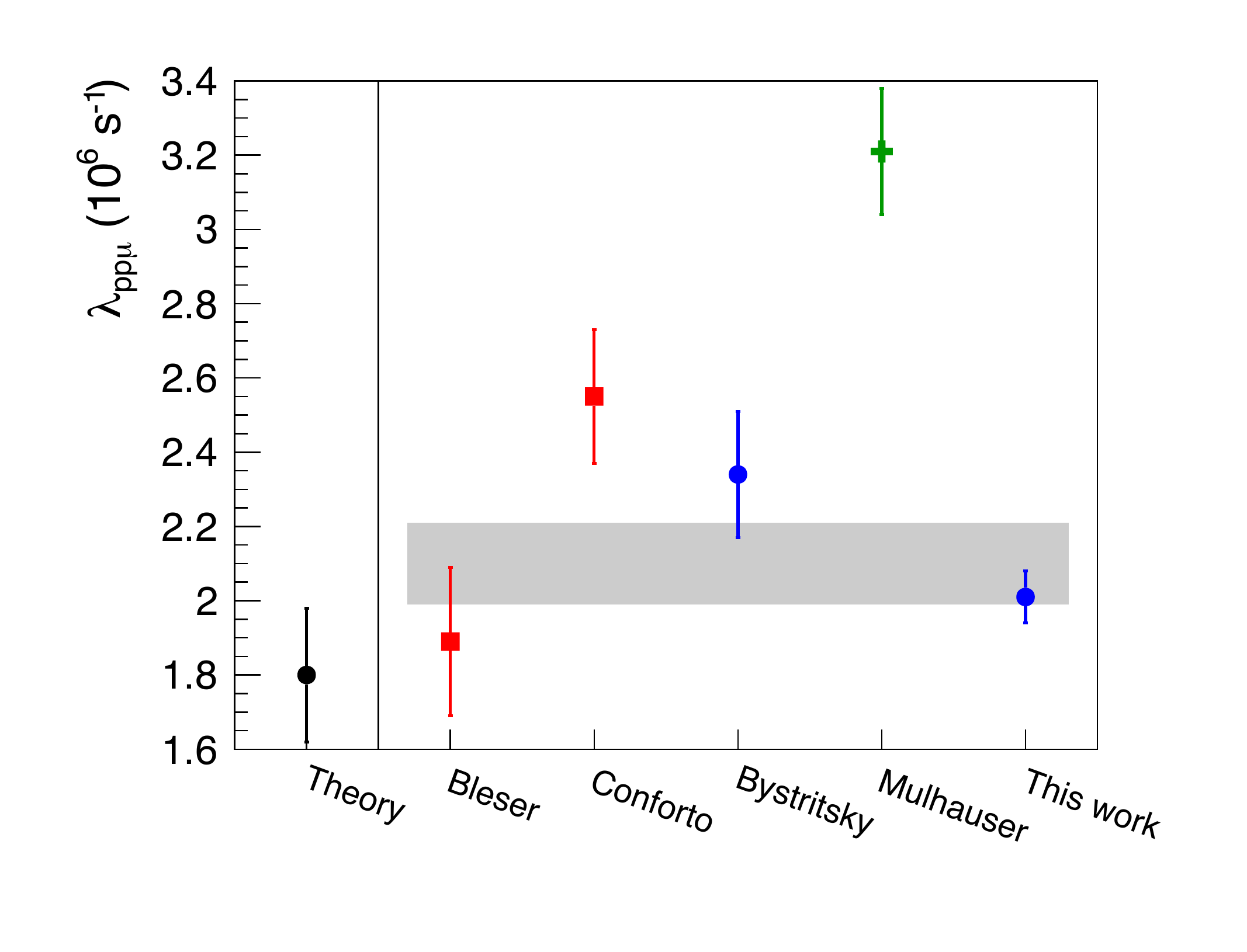}}
  \caption{ (Color online) Comparison of a theoretical
            calculation~\cite{Faifman:1989} and experimental
            measurements of the molecular formation rate $\lppmu$.
            Red squares denote liquid-hydrogen targets (Bleser et
            al.~\cite{PhysRev.132.2679}, Conforto et
            al.~\cite{Conforto:1964}), the green cross denotes a solid-target
            measurement (Mulhauser et al.~\cite{Mulhauser:1996zz}),
            and the blue circles denote two measurements in a
            gaseous hydrogen environment (Bystritsky et
            al.~\cite{Bystritskii:1976} and this paper). The shaded region
            corresponds to an updated world average of the experimental
            results, excluding the outlying solid-target data point.}
  \label{fig:ppmu}
\end{figure}

Figure ~\ref{fig:ppmu} plots the relevant experimental and theoretical
determinations of $\lppmu$, including that presented in this paper. 
The experimental data are not completely consistent. 
The higher $\lppmu$ value measured in the solid-target experiment 
could originate from comparatively slower thermalization of the $p\mu$ 
atoms via elastic collisions with the solid hydrogen lattice~\cite{Adamczak:2007}. 
Review~\cite{Kammel:2010zz} excluded the solid-hydrogen result to
obtain the experimental world average $\lppmu = (2.3 \pm 0.2) \times 10^6$\,\ivs,
where the uncertainty has been inflated to account for the inconsistencies among
the contributing measurements.

\subsection{Impact of Molecular Effects on Muon Capture Experiments}
Muon capture experiments determine $\LS$ either by measuring the rate of neutron
emission according to Eq.~\eqref{eq:Nc} (``neutron method'') or by inferring the muon disappearance
rate in hydrogen, $\lmum$, from the time distribution of electrons, Eq.~\eqref{eq:Ne} (``lifetime method''). 
While the neutron method does not require high statistics, its precision is fundamentally limited
by the fact that the neutron detection efficiency $\epsilon_c$ must be known to a level that is difficult 
to achieve in practice. Conversely, the lifetime method requires high statistics but absolute detection 
efficiency is not a factor. The basic idea of the lifetime method can be illustrated by considering the ideal case in which only
the $p\mu$ state is populated. In that case the electron time distribution Eq.~\eqref{eq:N} simplifies to 
\begin{equation}
N_e(t) \propto e^{-(\lmup +\LS) t}=e^{-\lmum t}
\label{eq:simple}
\end{equation} 
and $\LS$ can be determined from the difference $\lmum-\lmup$. 

In reality, experiments must always account for effects arising from the existence of muonic molecules. 
The lifetime method was pioneered by an experiment at Saclay~\cite{Bardin:1980mi} which used an LH$_2$ 
target ($\phi=1$); the full kinetics of Eq.~\eqref{eq:N} therefore needed to be considered, and this led to significant 
uncertainty in the interpretation of the experiment's results. The MuCap experiment~\cite{Andreev:2012fj} used a 
low-density hydrogen target ($\phi$=0.01) in order to more closely approach the ideal case of a purely $p\mu$ 
system. In the following we analyze the impact of muon chemistry on the lifetime method only; the reader is referred 
to review~\cite{Kammel:2010zz} for a more comprehensive treatment of muon capture experiments in hydrogen.

\begin{figure}[htb]
{\centering \includegraphics[width=0.99\linewidth]{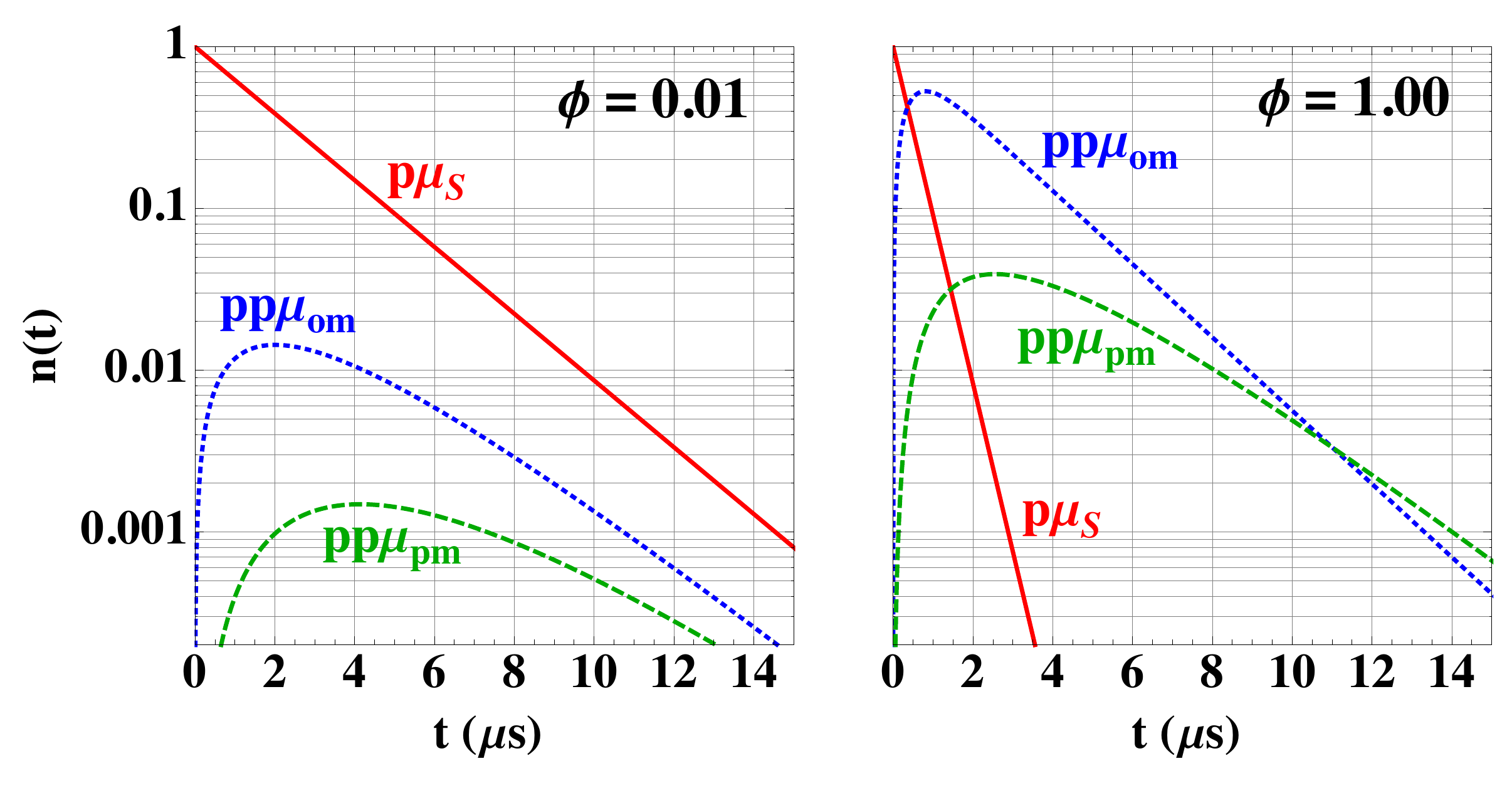}}
 \caption{(Color online) Calculated muonic-state populations for
            (left)~the hydrogen density in the MuCap experiment,
            $\phi$=0.01, and (right) LH$_2$, $\phi$=1.
            In MuCap 97\% of all captures proceeded from the $p\mu$
            singlet state, while in LH$_2$ capture takes place
            predominantly from $pp\mu$ molecules.}
    \label{fig:Times}
\end{figure}

Figure~\ref{fig:Times} shows the time distributions of $p\mu$ and $pp\mu$
populations in the hydrogen targets used in the MuCap~\cite{Andreev:2012fj}
and Saclay~\cite{Bardin:1980mi} experiments. 
At the lower target density used in MuCap, muons remain predominantly
in the singlet $p\mu$ state over the course of the typical measurement
period of 15~microseconds. There is nevertheless non-negligible 
formation of \ppmuO\ molecules, and therefore good knowledge of the rate 
$\lppmu$ of the process is necessary for correct interpretation of the experiment.
In contrast, in the LH$_2$ target used in the Saclay experiment the muon
quickly populates the \ppmuO\ state, within 1\,$\mu$s, and the subsequent
depopulation of the \ppmuO\ state to the \ppmuP\ state at rate $\lop$ is the 
crucial element to interpreting the experiment.

\section{Experimental Method}
\label{MuCap}
\subsection{MuCap Apparatus}

\begin{figure}[hbt]
{\centering    \includegraphics[width=0.99\linewidth]{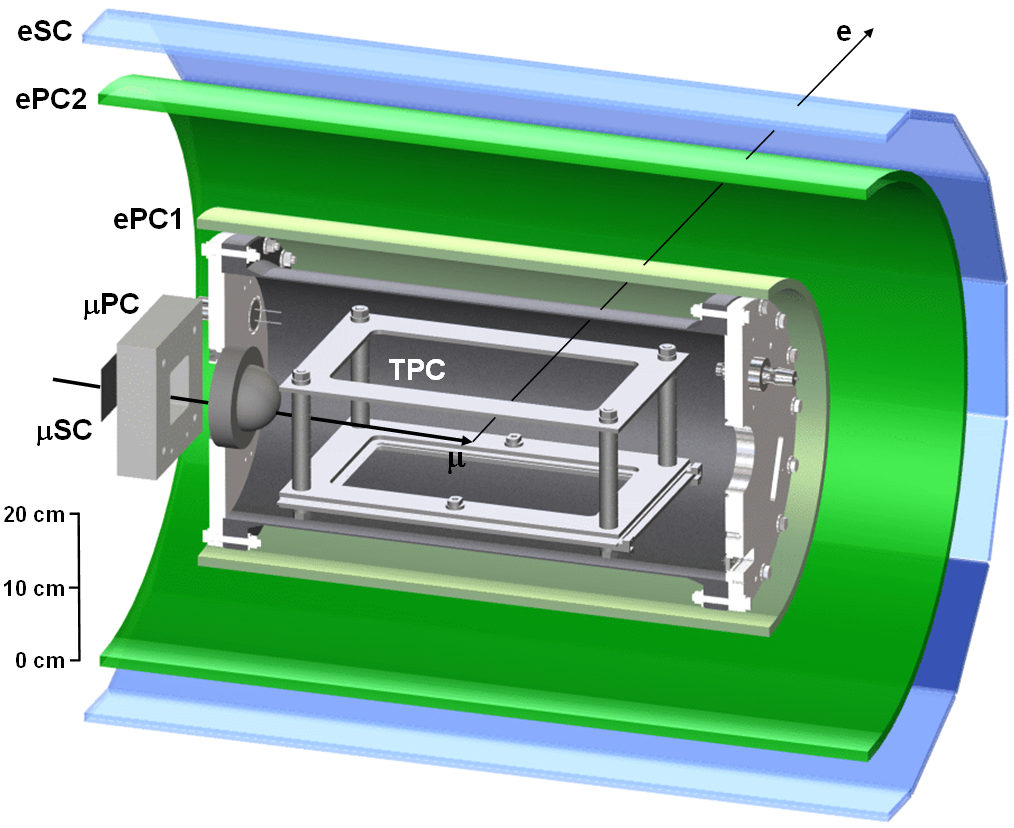}}  
    \caption{Simplified cross-sectional view of the MuCap detector setup. Neutron detectors not shown. The main components are described in the text.(Figure reproduced from \cite{Andreev:2007wg}.)}
    \label{fig:mucapdetector}
\end{figure}

The MuCap detector (Fig.~\ref{fig:mucapdetector}) will be described here only in brief; greater detail is available in~\cite{Andreev:2012fj, Andreev:2007wg, Kiburg:2011, Knaack:2012}. The experiment was located at the $\pi E3$ secondary muon beamline of the 590~MeV proton cyclotron at the Paul Scherrer Institute. Low-energy muons (34~MeV/$c$) passed through a scintillator counter ($\mu$SC) and a wire-chamber plane ($\mu$PC) before coming to a stop inside a 10-bar hydrogen time projection chamber~(TPC). 

The $\mu$SC provided the start signal for the muon lifetime measurement, and the $\mu$SC and $\mu$PC together provided efficient pileup rejection which enabled selection of events in which only a single muon was present in the TPC. The~TPC \cite{Egger:2011zz, Egger:2014bua} provided tracking of incoming muons and clear identification of each muon's stopping location by detecting the large peak in energy deposition at the end of the muon's Bragg curve. The trajectories of outgoing decay electrons were reconstructed by two concentric multi-wire proportional chambers~(ePC1 \& ePC2), while a scintillator barrel~(eSC) provided the stop time for the lifetime measurement.

Fiducial cuts can be applied to the TPC data to select muons that stopped in the hydrogen gas, far away from any rate-distorting $\rm{Z}>1$ materials. The three-dimensional electron tracking makes it possible to correlate a decay electron with the stopping point of its parent muon, thereby increasing the signal-to-background ratio (see \cite{Kiburg:2011, Knaack:2012} for details).

\subsection{ Measurement of $\lppmu$ in Argon-doped Hydrogen}

For this measurement we introduced argon to the otherwise ultra pure
hydrogen gas, which was at density $\den=0.0115 \pm 0.0001$. 
The atomic concentration of argon was $c^{}_{\rm{Ar}}=19.6 \pm 1.1$~ppm,
as measured both volumetrically during the initial filling and by gas chromatography at the end of the measurement. 
The two concentration measurements were consistent, but we have conservatively expanded the uncertainty to cover the
uncertainties of both. The gas density has been derived from the temperature and pressure, which were continuously monitored.

In principle, two time distributions, $N_c(t)$ (Eq.~\eqref{eq:Nc}), and
$N_e(t)$ (Eq.~\eqref{eq:Ne}), are experimentally observable. 
The former can be measured either by using the TPC to detect nuclear recoil signals 
from $\mu + \rm{Ar} \rightarrow \rm{Cl}^* + \nu$ capture events, or by using liquid 
scintillators to detect neutrons emitted by the excited final-state nucleus.
There are two disadvantages to measuring $N_c(t)$. First, the spectrum determines 
$\Gpmu$ (c.f.~$\nmuZ(t)$ in Eqs.~\eqref{eq:fullsolutions}) and therefore only the sum of the two 
transfer rates $\Lppmu$ and $\LpAr$, not the individual rates themselves, and
$\LpAr$ is not known with sufficient precision to enable $\Lppmu$ to be extracted independently. 
Second, there are significant systematic uncertainties relating to spatial pileup of TPC signals
from the stopping muon and the capture recoil, and to uncertainties in the neutron time of flight.

The MuCap experiment was designed to detect decay electrons, so we used a high-statistics sample of $N_e(t)$ to extract $\lppmu$.
If a muon decays it cannot undergo nuclear capture, eliminating the possibility of distortions in muon stop identification due to 
additional energy deposit from capture recoils. 
Consequently, the analysis and systematic uncertainties were very similar to those developed for the
earlier lifetime experiment measuring $\LS$~\cite{Andreev:2012fj}. 

The decay-electron analysis works as follows. With the judicious choice of argon
concentration $c^{}_{\rm{Ar}}=\mathcal{O}(20\,\rm{ppm})$, the disappearance rates $\Gpmu$ and $\GAr$ in Eq.~\eqref{eq:eigen} are sufficiently different as to allow them to be unambiguously extracted from a fit to the corresponding decay-electron time spectrum. 
The argon capture rate $\LAr$~\cite{Suzuki:1987jf} is three times higher than the muon decay rate and therefore transferred muons disappear quickly.  Under our conditions, the contributions of $\Lppmu$ and $\LpAr$ to the total \pmuS\ disappearance rate $\Gpmu$
were 4\% and 8\%, respectively. As above, the eigenvalue $\Gpmu$ alone would only determine the sum of two unknowns, $\Lppmu$ and $\LpAr$. However, both rates enter into the coefficients $c_{i}^\alpha$ in Eq.~\eqref{eq:alpha} in independent combinations, as can be seen from the full solutions in the appendix. A combined fit can therefore simultaneously determine $\Lppmu$, $\LpAr$, and, as a
byproduct, $\LAr$, without any need for absolute normalization. To address concerns about the uniqueness and stability
of this multi-parameter fit to a single distribution, we performed extensive pseudo-data
Monte Carlo studies of the full kinetics equations; good convergence was observed.

\section{Analysis and Results}
\label{Experiment_ppm}

\subsection{Data Analysis\label{sec:systematic}}

A total of $7.2\times10^8$ fully reconstructed muon decay events
were used in the present analysis. These events were selected via application 
of our standard cuts, described in Ref.~\cite{Andreev:2012fj}. 
Each event was required to involve a pileup-free muon stop in the TPC fiducial 
volume, $\Delta x \times \Delta y \times \Delta z =
10.4\times8.0\times20.4~\textrm{cm}^{3}$.
The decay-electron trajectories were reconstructed from 
spatial and temporal coincidences among the two cylindrical wire chambers 
and the two layers of plastic scintillators. 
Once the set of good events had been selected, the time differences between the
fast signals of the electron scintillator eSC and the muon beam scintillator 
$\mu$SC were histogrammed and the resulting decay time spectrum
was fitted with the function 
\begin{equation}
N(t) = A \left[\nmup(t)+\npara(t)+\northo(t)+\epsilon\; h\; \nmuZ(t)\right] + B 
\label{eq:fitfun}
\end{equation}
using the MINOS package.
This fit function is identical to Eq.~\eqref{eq:Ne} apart from the
introduction of a flat background term B. The relative efficiency
$\epsilon$ is defined as $\epsilon \equiv \epsilon'_e/\epsilon_e$.

To accommodate the time dependence of $\lambda_{p\rm{Ar}}$ in a nearly
model-independent way, this rate was parametrized in the form
\begin{equation}
\lambda ^{\rm{fit}}_{p\rm{Ar}}(t)=\lambda_{p\rm{Ar}} (1-\alpha e^{-\beta t}),
\label{eq:lpAr}
\end{equation}
where $\alpha$ and $\beta$ were extracted from Fig.1 in
\cite{PhysRevA.55.3447}. The parameter $\beta$ characterizes
$p\mu$ thermalization and was scaled down by~1.5 from
the value in~\cite{PhysRevA.55.3447}, as that experiment used a 15-bar target 
whereas MuCap used a 10-bar target. The scaling of $\alpha$ with pressure depends
on the initial population of hot $p\mu$ atoms after the muonic
cascade, which, according to theory~\cite{Markushin:1994}, should increase by $\sim$10\%
with a pressure increase from 10 to 15 bar. 
We did not change the value of 
$\alpha$ extracted from~\cite{PhysRevA.55.3447},
but we assigned it a conservative 50\% uncertainty.
The final fit method used numerical integration with the values
listed in Table~\ref{tab:fixedparameters}. The analytical
solution~\eqref{eq:ANe} was used for cross checks.

The fitting procedure using Eq. \eqref{eq:fitfun} requires a timing calibration to assert that the muon arrival time is at $t=0$. For that, the rising edge of the histogrammed differences of the $\mu SC$ and the sixteen eSC subdetectors were fitted individually. This determined timing calibration offsets for each eSC detector with a precision of 2\,ns. The sixteen offsets were then applied to their corresponding spectrum before the sum of all time distributions was fit with Eq. \eqref{eq:fitfun}.

\begin{table}[htb]
\caption{Experiment specific parameters used in
  the fit of Eq.~\eqref{eq:fitfun} to the
  data. See text for details on their evaluation.
  \label{tab:fixedparameters}}
{\centering \begin{tabular}{ll}
\hline
\hline
\T \B Parameter & Value \\
\hline
\T $c^{}_{\rm{Ar}}$ &$19.6 \pm 1.1$ ppm\\ 
$\phi$ & $0.0115 \pm 0.0001$ \\
$f$ & $5\pm 1\times 10^{-4}$ \\
$\epsilon$ & $0.996 \pm 0.003$\\
$c^{}_{\rm{O}} $& $57 \pm 57$ ppb  \\
$c^{}_{\rm{N}} $& $115 \pm 115$ ppb  \\
$\alpha$ & $0.25 \pm 0.12$   \\
\B $\beta$ & $1.0 \pm 0.2 \times 10^7 ~\mathis$   \\
\hline  
\hline
\end{tabular}}
\end{table}
The fit was performed over the range $[0.12 \mu\rm{s}, 20 \mu\rm{s}]$. 
Five quantities were treated as free parameters: $\Lppmu$, $\LpAr$, $\LAr$,
the normalization $A$, and the background term $B$. 
All other parameters were fixed in the fit to the values in
Table~\ref{tab:relevantrates} and,  for experiment specific 
parameters, according to the values given in Table~\ref{tab:fixedparameters}. 
The initial $\mu\rm{Ar}$ formation fraction $f=(5 \pm 1)\times 10^{-4}$ is the sum of two components, $f_c$ and $f_e$. The atomic capture ratio for argon relative to hydrogen has been measured to be $f_c=(9.5\pm 1.0 )c_{\rm{Ar}} =(1.87 \pm 0.20) \times 10^{-4}$~\cite{PhysRevA.56.468}. An additional initial population
$f_e=(1.66\pm 0.34) f_c$ from excited-$p\mu$-state transfer has been observed in a target at 15-bar pressure~\cite{PhysRevA.56.468}. We account for this by using $f_e= (3.1 \pm 0.9) \times 10^{-4}$, in which the uncertainty has been conservatively enlarged to accommodate the possibility of a pressure dependence.

The energy spectra of decay electrons emitted from $p\mu$ and Ar$\mu$ atoms 
are different, which leads to a difference in the corresponding detection efficiencies.
We used the energy spectrum
calculated in~\cite{PhysRev.124.904} and folded it together with the
energy-dependent detector efficiency obtained from a full Geant4
simulation. The resulting relative efficiency, 
$\epsilon=0.996 \pm 0.003$, shows that the thin layers of the
MuCap electron detectors are not very sensitive to spectral differences at
higher energies.

After the fit, small corrections were applied to the fitted rates to account for the presence of chemical impurities
oxygen and nitrogen, with atomic concentrations $c^{}_O$ and $c^{}_N$, respectively. 
This procedure is discussed in the next section.

\subsection{Results and Systematic Uncertainties}

\begin{figure}[htb]
{\centering\includegraphics[width=0.99\linewidth]{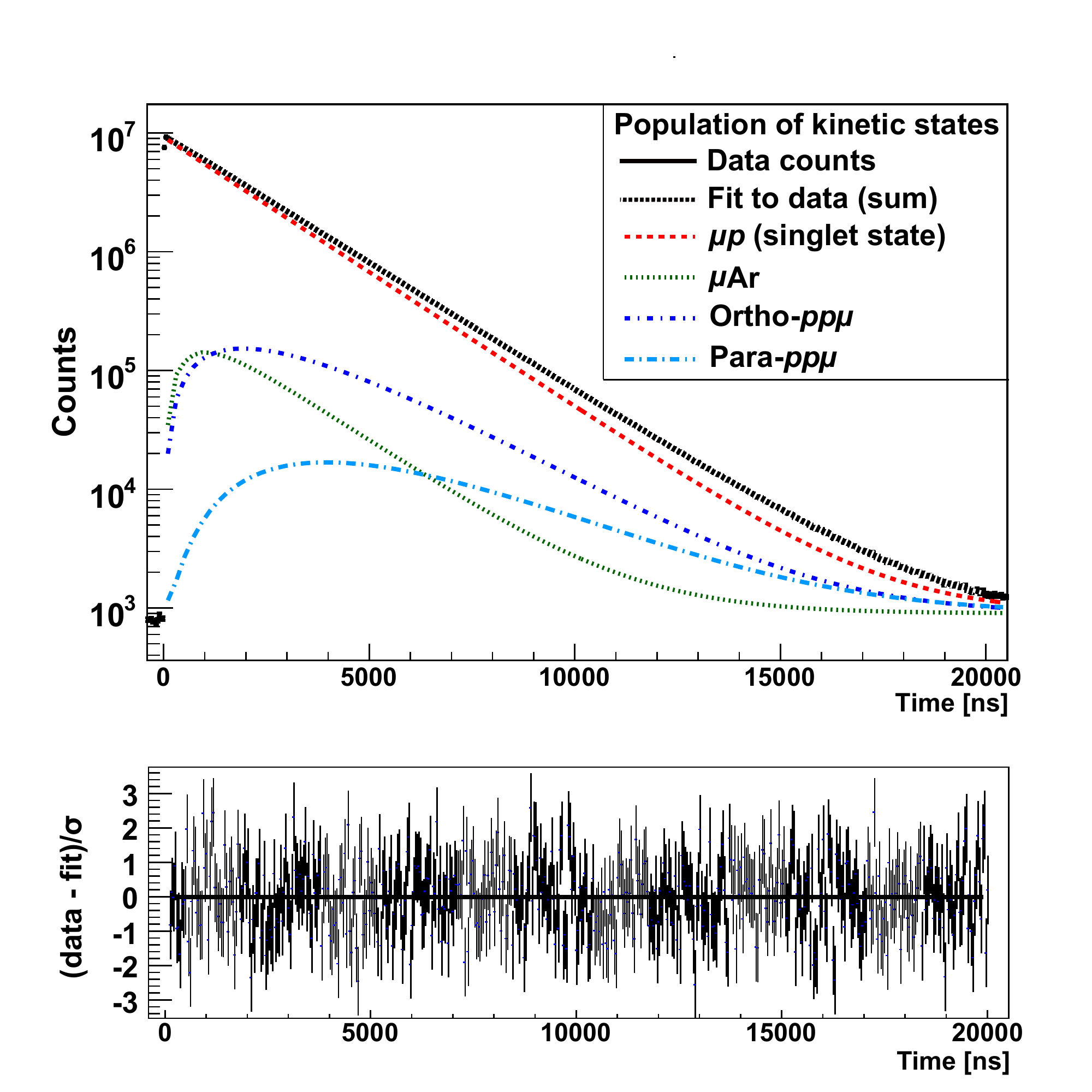}}
\caption{(Color online.) Upper Panel: Fit to the decay electron
  time spectrum using Eq.~\eqref{eq:fitfun}. The data is shown as the black line.
  The colored curves depict the time-dependent contributions from the
  kinetic states; the black dashed line is the fitted sum. Lower Panel: The
  normalized residuals between the data and the fit function, $(N_i - N(t_i))/\sigma_i$,
  indicate good agreement in the fitted range 0.12--20~$\mu s$.}
\label{FitHere}
\end{figure}

The fit to the data is plotted in Fig.~\ref{FitHere}. The upper
panel shows the decay electron time spectrum alongside 
the time distributions of the parent muon populations $\nmup{}(t)$,
$\nmuAr{}(t)$, $\northo{}(t)$, and $\npara{}(t)$ determined
by the fit. The lower panel displays the residuals,
i.e., the differences between the data and the fit function normalized
by the uncertainty of each data point. 
The good agreement between the data and the fit function is
demonstrated by the reduced $\chi^2/\rm{DOF} = 0.983 \pm 0.064$. 

Table~\ref{Results_ppm} presents the fit results for the three rates
$\Lppmu$, $\LpAr$, and $\LAr$. The table also lists systematic
corrections $\Delta$ and the systematic uncertainties $\delta$ 
resulting from a $\pm 1\sigma$ variation of the fixed parameters listed in
Tables~\ref{tab:relevantrates} and~\ref{tab:fixedparameters}. 

\begin{table}[htb]
\caption{Fit results for $\Lppmu$, $\LpAr$, and $\LAr$, as well as
their associated systematic corrections~($\Delta$) 
and uncertainties~($\delta$). The final error on each rate is the
quadrature sum of the contributing uncertainties.}
{\centering \begin{tabular}{lrr@{\ \ \ }rr@{\ \ \ }rr}
\hline
\hline
\T \B & \multicolumn{2}{c}{$\Lppmu{}$\,[\ivs{}]} & \multicolumn{2}{c}{$\LpAr{}$\,[\ivs{}]} & \multicolumn{2}{c}{$\LAr{}$\,[$10^{2}$\,\ivs{}]\vspace{0.5mm}}\\
\hline
\T \B Fit 	       & 22\,996 & $\pm 647$  & 43\,799 & $\pm 151$ & 13\,023 & $\pm 147$\\
\hline
\T \B Systematic & \rule{0cm}{3mm}$\Delta$ & $\delta$ & $\Delta$ & $\delta$ & $\Delta$ & $\delta$ \\
\hline
\T Timing calibration     & &  $39$  & & $13$  & & $35$\\
Efficiency $\epsilon$  & &  $37$   & & $ 13$  & & $ 34$\\
Huff factor $h$        & &  $45$   & & $ 15$  & & $ 27$\\             
$f$    		       & &  $46$   & & $ 12$  & & $ 39$\\
$\LS,~\LO,~\LP$        & &  $13$   & & $ 16$  & & $ 5$\\
$\lop{}$      	       & &  $31$   & & $ 3$   & & $ 2$\\ 
$\Lpf$	               & &  $9$    &	&      & & \\
Epithermal             & &  $329$  & & $85$   & & $278$ \\
\B H$_2$O and N$_2$      & 116 & $116$ & $-15$ & $ 15$  & & \\
\hline
\T \B Final result	       & 23\,112 & $\pm 741$  & 43\,784 & $\pm 177$ & 13\,023 & $\pm 322$\\
\hline
\hline
\end{tabular}}
\label{Results_ppm}
\end{table}

The fit did not explicitly model effects from the accumulation
of nitrogen and oxygen in the hydrogen due to outgassing from the TPC
vessel. Instead, a correction $\Delta$ was applied to the fitted values of 
both $\Lppmu$ and $\LpAr$. During its main run MuCap achieved hydrogen chemical
purity levels of better than 10 ppb, but during the argon-doped measurement
the TPC was disconnected from the hydrogen circulation and purification
system~\cite{Ganzha:2007uk}. After six days, atomic concentrations of 
$c^{}_{\rm{O}}$=115~ppb of oxygen (in the form of water vapor) 
and $c^{}_{\rm{N}}$=230~ppb of nitrogen were observed using
a humidity sensor and gas chromatography, respectively. Due to the
higher muon transfer and capture rates for oxygen compared to nitrogen,
transfer to oxygen is the dominant effect needing to be taken into
account in the correction to the measured rates. A series
of pseudo data were generated based on the kinetics in Eq.~\eqref{eq:N},
with transfer to and capture on chemical impurities included. 
MuCap had previously measured these rates
independently using impurity-doped hydrogen mixtures. Our
result for $\lambda_{p\rm{N}}$ agreed with previous measurements, but our results for $\lambda_{p\rm{O}}$
(measured via water doping) were nearly two times higher than the value quoted
in Table~\ref{tab:relevantrates}. For internal consistency we used the transfer
rates measured by MuCap in our simulation.
The pseudo data were then fitted with Eq.~\eqref{eq:fitfun} to extract the shifts in the rates $\Lppmu$, $\LpAr$, and $\LAr$ as a function of the oxygen concentration $c^{}_{\rm{O}}$. As the exact time dependence of the impurity buildup was unknown, conservative estimates of $c^{}_{\rm{O}}=57 \pm 57$~ppb and $c^{}_{\rm{N}}=115 \pm 115$~ppb were used to cover all possible accumulation scenarios. The impurity-related corrections to $\Lppmu$ and $\LpAr$ were determined to be $\Delta^{}_{pp\mu} = 116 \pm 116$\,\ivs\ and $\Delta^{}_{p\rm{Ar}} = -15 \pm 15$\,\ivs.

The final results for the fitted rates after applying the impurity-related corrections
and summing all systematic uncertainties (Table~\ref{Results_ppm}) are
\begin{eqnarray}
\Lppmu&=&2.311 \pm 0.074  \times 10^4\;\rm{s}^{-1}\nonumber\\
\LpAr   &=&4.378 \pm 0.018   \times 10^4 \;\rm{s}^{-1}\label{eq:FinalFit}\\
\LAr     &=&1.302 \pm 0.032   \times 10^6 \;\rm{s}^{-1} ~.\nonumber
\end{eqnarray}
From these one can deduce the normalized rates
\begin{eqnarray}
\lppmu&=&2.01  \pm  0.07  \times 10^6\;\rm{s}^{-1} \nonumber\\
 \lpAr &=&1.94 \pm 0.11 \times 10^{11} \;\rm{s}^{-1}
\label{eq:FinalFitNormalized}
\end{eqnarray}
using Eqs.~\eqref{eq:effectiverate} and \eqref{eq:ImpurityTransfer}, respectively.

The normalized correlations among the five free fit parameters are presented in Table~\ref{ElCorr}. These correlations are incorporated into the uncertainties on the final results.

\begin{table}[ht]
\caption{Normalized correlation coefficients of the free parameters in the 
fit to the decay electron time spectrum.}
{\centering \begin{tabular}{ccccc}
\hline
\hline
\T \B Rates & $\Lppmu{}$ & $\LpAr{}$ & $\LAr$ & A\\
\hline
\T $\LpAr{}$ & 0.9548 &  &  & \\
$\LAr$ & -0.8021 & -0.9011 &  & \\
A & 0.0495 & 0.0269 & 0.0234 & \\\
\B B & -0.6603 & -0.5479 & 0.4189 & -0.1082\\
\hline
\hline
\end{tabular}}
\label{ElCorr}
\end{table}

Our result for $\Lppmu$ is about $1\sigma$ larger than
the value we obtained in~\cite{Andreev:2012fj} due to the more refined analysis
in this paper and the correction of a numerical error in the fitting
code. 
As regards the transfer rate to argon $\lpAr$, there is a wide spread
of experimental results obtained with different methods and target conditions, clustered around $1.4 \times 10^{11}$\,\ivs, $3.6 \times 10^{11}$\,\ivs, and $9 \times 10^{11}$\,\ivs,
as discussed in \cite{PhysRevA.55.3447}. Our value $\lpAr= 1.94 \pm
0.11 \times 10^{11}$\,\ivs\ is close to the most recently published value, $1.63\pm 0.09
\times 10^{11}$\,\ivs~\cite{PhysRevA.55.3447}, albeit $2.2\sigma$ higher.
Note that the uncertainty in
the argon concentration only enters into the extraction of the normalized rate $\lpAr$, 
while in the fit to determine $\lppmu$ effective rates are being used
which are independent of $c_{\rm{Ar}}$. Our result for the muon's nuclear capture rate on argon, 
$\LAr$, agrees well with the values in the literature, $1.20 \pm 0.08
\times 10^{6}$\,\ivs~\cite{Bertin1973} and $1.41 \pm 0.11 \times 10^{6}$\,\ivs~\cite{Carboni1980} .

\subsection{Consistency Checks}

\begin{figure}[htb]
{\centering \includegraphics[width=1\linewidth]{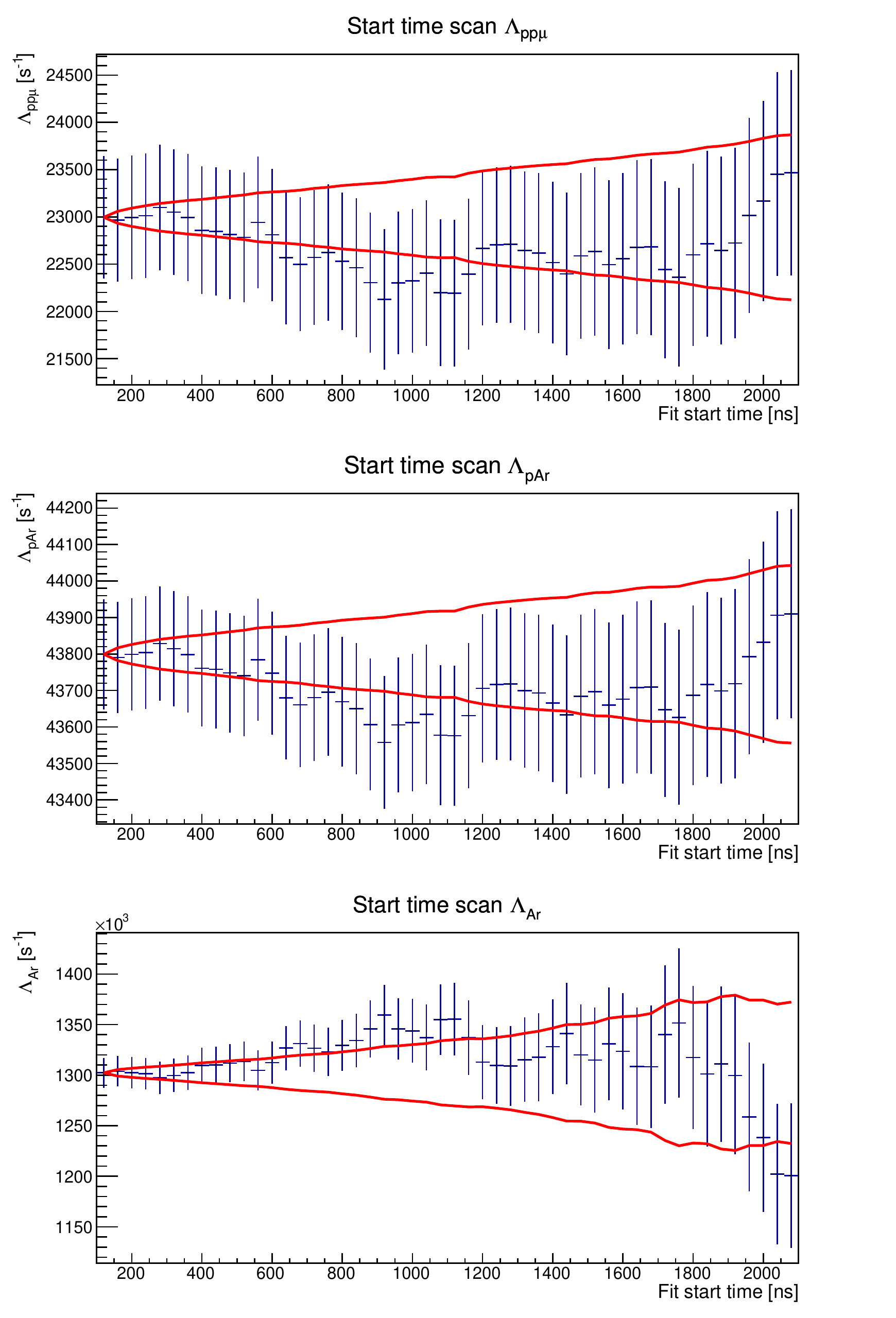}}
	\caption{(Color online) Results from a scan over the fit start time. 
		Each plot shows the fit results (blue points) for a particular rate as the start time is varied. 
          The variation of each rate is consistent with the expectations from the one-sigma statistically allowed set-subset deviation (red line).}
\label{ElStartTime}
\end{figure}

The fit start time was varied to check for any distortions or physical effects not accounted for by the fit function. Figure~\ref{ElStartTime} shows the progressions of the fitted rates as the fit start time was increased in steps from its standard value of $0.12~\mu\rm{s}$.
The red lines denote the $\pm1\sigma$ variation allowed because of the set-subset statistics involved in this procedure. Each rate is statistically self-consistent across the fit start time scan.

In the fit to the data using Eqs.~\eqref{eq:Ne} and \eqref{eq:fitfun},
the capture rate $\LS$ is required as an input to extract $\lppmu$, while the latter
is itself used in the determination of $\LS$. 
This interdependency is not a
problem because all fitted rates in
Eq.~\eqref{eq:FinalFit} depend only very weakly on the hydrogen capture rates,
as quantified in
Table~\ref{Results_ppm}. We explicitly iterated the procedure
(obtaining fit results with $\LS$ as input, using the results to correct $\LS$, repeating 
with the adjusted $\LS$) to arrive at a stable, self-consistent solution, and we found that the 
results for $\Lppmu$, $\LpAr$, and $\LAr$ were changed by less than one tenth of their uncertainties.

Lastly, the reproducibility of the fit was tested by generating $10^4$ pseudo-data histograms using the final fit parameters in Eqs.~\eqref{eq:FinalFit}, and fitting each pseudo experiment in the same manner as the real data.
The fits consistently yielded the input values, and the simulated data reproduced the same fit uncertainties listed in Table~\ref{Results_ppm}.

\section{Relevance to the Interpretation of the MuCap Experiment}
\label{Interpretation}

In Section~\ref{CaptureChemistry} the influence of the molecular rates $\lppmu$ and $\lop$ on muon kinetics in hydrogen was described. The MuCap experiment measured the effective muon disappearance rate $\lmum$ in low-density ultra-pure hydrogen by fitting the observed decay electron time distribution with a three-parameter function, $f(t) = A e^{-\lmum t} + B$. Taking $pp\mu$ formation into account, the disappearance rate can be expressed as
\begin{equation}
 \lmum= \lmup + \Delta\lambda^{}_{p\mu} + \LS + \Delta\Lambda^{}_{pp\mu} ~.
\label{eq:decayrate}
\end{equation}
Here $\Delta\lambda^{}_{p\mu} = -12.3$\,\ivs\ is a calculable bound-state modification to the muon decay rate in the $p\mu$ 
system~\cite{Uberall:1960zz, VonBaeyer:1979yg}, and $\Delta\Lambda^{}_{pp\mu}$ is a modification to $\LS$ accounting
for the small population of muonic molecules and the fact they have unique capture rates.
In the following we summarize the derivation of $\Delta\Lambda^{}_{pp\mu}$, based on our improved measurement of 
$\lppmu$ at conditions nearly identical to those of the main MuCap experiment. 

The derivation is based on high-statistics simulations of the full kinetics described by Eqs.~\eqref{eq:fullsolutions}. 
Since the MuCap measurement of $\LS$ was performed using pure hydrogen gas, for the simulations the Z channel
was used to model the small amounts (few ppb) of oxygen and nitrogen impurities that were observed to have 
outgassed from the hydrogen vessel's walls. The relevant input parameters for the simulation were those in 
Tables~\ref{tab:relevantrates} and~\ref{tab:fixedparameters}.
An accidental background was added to make the signal-to-background level commensurate with that in the MuCap data. 
Time distributions of $10^{12}$ decay electrons were generated for two different cases: 
$\lppmu=0$ and $\lambda^{\rm{MuCap}}_{pp\mu}$. 
The previous MuCap analysis~\cite{Andreev:2012fj} was performed using the preliminary value 
$\lambda^{\rm{MuCap}}_{pp\mu}=  1.94 \pm  0.06  \times 10^6\;\rm{s}^{-1}$; here we update the analysis using our new
result in Eq~\eqref{eq:lppm_MuCap}. To determine the effect on the MuCap result for $\LS$, 
we fit the simulated time distributions with the same three-parameter function used to fit the data. 
The relevant correction is then obtained via
\begin{equation}
\Delta \Lambda^{}_{pp\mu} = \lmum(\lambda^{\rm{MuCap}}_{pp\mu}) - \lmum(\lppmu=0) ~,
\end{equation}
where the $\lmum$ values are obtained from fits to the two simulated data sets generated using different $\lppmu$ values. 
The uncertainty in $\Delta \Lambda^{}_{pp\mu}$ is estimated in a similar manner, by generating 
pseudo data while varying the parameters entering the kinetic equations by $\pm 1 \sigma$ individually. 
The resulting fit determines the final correction for the MuCap experiment to be 
$\Delta \Lambda^{}_{pp\mu}=-18.4 \pm 1.9 ~\rm{s}^{-1}$, 
which is smaller than the correction in~\cite{Andreev:2012fj} by $0.7$~s$^{-1}$. 
Thus the updated value of $\lambda^{\rm{MuCap}}_{pp\mu}$ induces a small shift of the singlet $p\mu$ capture rate measured by MuCap from 
$\LS=714.9 \pm 5.4_{\mathrm{stat}} \pm 5.1_{\mathrm{syst}} ~\mathis$ obtained in \cite{Andreev:2012fj} to
\begin{equation}
\LS=715.6 \pm 5.4_{\mathrm{stat}} \pm 5.1_{\mathrm{syst}} ~\text{s}^{-1} ~.
\end{equation}
The value of the pseudoscalar coupling constant, $\gpm = 8.06 \pm 0.55$ extracted in \cite{Andreev:2012fj}, is
correspondingly changed by $-0.045$, i.e. by only 8\% of its uncertainty.
   
From our simulations we can determine the dependence of $\Delta \Lambda^{}_{pp\mu}$
on molecular parameters,
\begin{equation}\label{eq:DLppm}
  \Delta \Lambda^{}_{pp\mu}=-18.4 \; [1 + a  (\lppmu-\lambda^{\rm{MuCap}}_{pp\mu}) + b (\lop-\lambda_{\rm{op}}^0)] ~,
\end{equation}
where $\lambda_{\rm{op}}^0$ is given in Table~\ref{tab:relevantrates}, $a = 4.7\times 10^{-7}$, and $b= 2.9\times10^{-6}$.
Using the new measurement presented in this paper, the total uncertainty in the MuCap capture rate $\LS$ 
due to $pp\mu$ formation is less than 2~\ivs\ and is dominated by $\lop$, while $\lppmu$ contributes only 0.6~\ivs.

\section{Summary}
\label{Summary}
The time spectrum of electrons emitted by the decay of muons stopped in argon-doped hydrogen were measured with the MuCap detector, for the purpose of determining the formation rate $\lppmu$ of $pp\mu$ muonic molecules. The TPC enabled selection of muons that stopped in the hydrogen, away from high-Z materials, and the electron tracker provided 3$\pi$ solid-angle coverage and enabled vertex matching with muon stops.  
We developed a detailed physics model to describe the time evolution of the atomic and molecular muonic states contributing to the decay electron spectrum, taking into account the energy dependence of the muon transfer rate $\lpAr$ from hydrogen to argon. 
We extracted $\lppmu$, $\lpAr$, and the muon capture rate in argon, $\LAr$, from a single fit to the decay electron time spectrum. 
Our results for $\lpAr$ and $\LAr$ agree with those from previous dedicated experiments. 
Our result for the $pp\mu$ formation rate,
\begin{equation}
\lambda^{\rm{MuCap}}_{pp\mu}= 2.01 \pm 0.06_{\mathrm{stat}}\pm 0.03_{\mathrm{syst}}\times10^6~\mathrm{s}^{-1},
\label{eq:lppm_MuCap}
\end{equation}
is 2.5~times more precise than previous measurements, which were performed under a variety of different experimental conditions and whose results disagreed beyond their uncertainties.
To obtain a new world average we used the procedures for averaging and inflating uncertainties advocated by the Particle Data Group~\cite{Agashe:2014kda}: we have included only the gas- and liquid-target experiments~\cite{Bystritskii:1976,PhysRev.132.2679,Conforto:1964}, choosing to omit the lone solid-target experiment~\cite{Mulhauser:1996zz} because of possible solid-state effects which are not well understood. The updated experiment world average then becomes
\begin{equation}
\lambda^{\rm{avg}}_{pp\mu} = 2.10 \pm 0.11\times10^6~\mathrm{s}^{-1} ~.
\end{equation}

The rate $\lambda_{pp\mu}$ was a necessary input to the MuCap experiment's recent precision determination 
of the nuclear capture rate on the proton, $\LS$~\cite{Andreev:2012fj}. MuCap was designed so 
that the majority of muons underwent capture in muonic $p\mu$ atoms, and formation of $pp\mu$ 
molecules changed the observed capture rate by only 2.5\%. 
However, given the inconsistency between existing $\lambda_{pp\mu}$ results it was difficult
to confidently estimate the uncertainty on the correction to $\LS$ for $pp\mu$ effects. 
Our new result for $\lppmu$, obtained at the same hydrogen density and temperature as in the main MuCap
experiment, leads to a well-defined correction to $\LS$, and the corresponding contribution to the total error is now minor.
The value of $\lambda^{\rm{MuCap}}_{pp\mu}$ presented here differs only slightly from the value 
used in~\cite{Andreev:2012fj}, and consequently the updated values for $\LS$ and the pseudoscalar
coupling $\gpm$ agree to better than $0.1\sigma$ with the values in that publication.

\section*{Acknowledgments}
We are grateful to the technical and scientific staff of the collaborating institutions, in particular the host laboratory the Paul Scherrer Institute, for their contributions. We thank J.D. Phillips for the GEANT4 simulation used in the determination of the electron efficiency.
This material is based upon work supported by the U.S.\ National Science Foundation; the U.S.\ Department of Energy Office of Science, Office of Nuclear Physics, under Award Number DE-FG02-97ER41020; the CRDF; the Paul Scherrer Institute; the Russian Academy of Sciences; and the Grant of the President of the Russian Federation (NSH-3057.2006.2). 
This work used the Extreme Science and Engineering Discovery Environment (XSEDE), which is supported by National Science Foundation grant number ACI-1053575.

\\

\appendix
\section{Solutions to the muon kinetics equations\label{sec:appendix}}
The differential equations in Eq. \eqref{eq:N} can be solved by determining the eigenvalues and
eigenvectors of the system. The time-dependent populations of the four muonic states are given by

\begin{widetext}
\begin{equation}
\begin{split}
\nmup(t) &= (1-f)\cdot\,e^{-\Gpmu t}_{}\\
\northo(t) &= (1-f)\cdot\frac{\Lof}{\Gom - \Gpmu}\cdot \left(e^{-\Gpmu t}_{} - e^{-\Gom t}_{}\right)\\
\npara(t) &= \frac{1-f}{\Gom-\Gpmu} \cdot \left(\frac{\Lof\lop}{\Gom - \Gpm} \cdot \big(e^{-\Gom t}_{} - e^{-\Gpm t}_{} ) + \frac{\Gpmu\Lpf - \Gom\Lpf-\Lof\lop}{\Gpmu - \Gpm} \cdot \big(e^{-\Gpmu t}_{} - e^{-\Gpm t}_{}\big)\right)\\
\nmuZ(t) &= (1-f)  \cdot \frac{\LpZ}{\Gz - \Gpmu} \cdot (e^{-\Gpmu t} - e^{-\Gz t}) + f \cdot e^{-\Gz t} ~.
\end{split}
\label{eq:fullsolutions}
\end{equation}
\end{widetext}
One simplistic but heuristic approximation is to 
neglect the small parameters $\LS$ and $\LP$ in the disappearance rates in Eqs.~\eqref{eq:eigen} and the initial
Z$\mu$ population $f$, and assume that $\Gz$ is large compared to all
other eigenvalues in Eq.~\eqref{eq:eigen}. 
In this limit the observable electron distribution, Eq.~\eqref{eq:Ne}, attains the simple form
\begin{equation}
N_e(t) \propto e^{-\lmup t} \left[ 1+ \frac{\LpZ}{\Lppmu} e^{-(\Lppmu + \LpZ) t} \right] ~,
\label{eq:ANe}
\end{equation}
which elucidates our strategy of determining $\Lppmu$ in a single fit.

\clearpage
\bibliography{MuonChemistry}

\end{document}